\documentclass[twocolumn,prb,amsmath,amssymb,floatfix,superscriptaddress,showpacs]{revtex4-1}
\usepackage{graphicx}
\usepackage{dcolumn}
\usepackage{bm}
\usepackage{xspace}
\begin{document}

\title{Enhanced low-energy magnetic excitations via suppression of the itinerancy in Fe$_{0.98-z}$Cu$_z$Te$_{0.5}$Se$_{0.5}$}
\author{Jinsheng Wen}
\altaffiliation{jwen@nju.edu.cn}
\affiliation{Center for Superconducting Physics and Materials, National Laboratory of Solid State Microstructures and Department of Physics, Nanjing University, Nanjing 210093, China}
\affiliation{Department of Physics, University of California, Berkeley, California 94720, USA.}
\affiliation{Materials Science Division, Lawrence Berkeley National Laboratory, Berkeley, California 94720, USA}
\author{Shichao Li}
\affiliation{Center for Superconducting Physics and Materials, National Laboratory of Solid State Microstructures and Department of Physics, Nanjing University, Nanjing 210093, China}
\author{Zhijun Xu}
\author{Cheng Zhang}
\affiliation{Condensed Matter Physics and Materials Science
Department, Brookhaven National Laboratory, Upton, New York 11973,
USA}
\author{M. Matsuda}
\affiliation{Quantum Condensed Matter Division, Oak Ridge National Laboratory, Oak Ridge,
Tennessee 37831, USA.}
\author{O.~Sobolev}
\author{J.~T.~Park}
\affiliation{Forschungsneutronenquelle Heinz Maier-Leibnitz (FRM-II), TU M\"{u}nchen, D-85747 Garching, Germany}
\author{A. D. Christianson}
\affiliation{Quantum Condensed Matter Division, Oak Ridge National Laboratory, Oak Ridge,
Tennessee 37831, USA.}
\author{E. Bourret-Courchesne}
\affiliation{Life Sciences Division, Lawrence Berkeley National Laboratory, Berkeley, California 94720, USA}
\author{Qiang~Li}
\author{Genda~Gu}
\affiliation{Condensed Matter Physics and Materials Science
Department, Brookhaven National Laboratory, Upton, New York 11973,
USA}
\author{Dung-Hai Lee}
\affiliation{Department of Physics, University of California, Berkeley, California 94720, USA.}
\affiliation{Materials Science Division, Lawrence Berkeley National Laboratory, Berkeley, California 94720, USA}
\author{J.~M.~Tranquada}
\affiliation{Condensed Matter Physics and Materials Science Department, Brookhaven National Laboratory, Upton, New York 11973, USA}
\author{Guangyong Xu}
\affiliation{Condensed Matter Physics and Materials Science
Department, Brookhaven National Laboratory, Upton, New York 11973,
USA}
\author{R. J. Birgeneau}
\affiliation{Department of Physics, University of California, Berkeley, California 94720, USA.}
\affiliation{Materials Science Division, Lawrence Berkeley National Laboratory, Berkeley, California 94720, USA}
\affiliation{Department of Materials Science and Engineering, University of California, Berkeley, California 94720, USA.}
\date{\today}

\begin{abstract}
We have performed resistivity and inelastic neutron scattering measurements on three samples of Fe$_{0.98-z}$Cu$_z$Te$_{0.5}$Se$_{0.5}$ with $z=0$, 0.02, and 0.1. It is found that with increasing Cu doping the sample's resistivity deviates progressively from that of a metal. However, in contrast to expectations that replacing Fe with Cu would suppress the magnetic correlations, the low-energy ($\leq 12$~meV) magnetic scattering is enhanced in strength, with greater spectral weight and longer dynamical spin-spin correlation lengths. Such enhancements can be a consequence of either enlarged local moments or a slowing down of the spin fluctuations. In either case, the localization of the conduction states induced by the Cu doping should play a critical role. Our results are not applicable to models that treat 3$d$ transition metal dopants simply as effective electron donors.  
\end{abstract}


\maketitle
\section{Introduction}
The effects of substitution of 3$d$ transition metals (such as Co, Ni, Cu, etc) on the crystal and magnetic structure, Fermi-surface topology, superconductivity, and magnetism in Fe-based superconductors have been widely discussed.~\cite{canfield:060501,PhysRevLett.110.107007,PhysRevLett.109.077001,PhysRevB.84.020509} Some initial studies on BaFe$_2$As$_2$ (Ba122) have suggested that 3$d$ metals such as Co partially substituted for Fe, act as effective electron donors.~\cite{canfield:060501,PhysRevB.82.024519,PhysRevB.84.020509} Such approaches typically describe the doping effects based on a rigid-band shift model.~\cite{PhysRevB.83.094522,PhysRevB.83.144512,JPSJ.80.123701}  However, such a model has faced serious challenges from both the experimental and theoretical perspectives,~\cite{PhysRevLett.105.157004,PhysRevLett.110.107007,2011arXiv1112.4858B,2011arXiv1107.0962B,PhysRevLett.109.077001} each of which indicates the inadequacy of a rigid-band description. Furthermore, there is a dichotomy between Co/Ni and Cu substitution effects in terms of the emergence of the superconductivity. In Ba122, with increasing Co or Ni doping, superconductivity appears concomitant with the suppression of the lattice distortion and magnetic order.~\cite{canfield:060501,PhysRevB.82.024519,PhysRevB.84.020509} However, superconductivity in the Cu-doped case has been observed only in one Cu concentration with $T_c\sim2$~K.~\cite{canfield:060501,PhysRevB.84.054540,PhysRevB.82.024519}

In this work, we explore the effects of transition metal substitution by carrying out resistivity and inelastic neutron scattering measurements on a different Fe-based superconductor system, namely Fe$_{1+y}$Te$_{1-x}$Se$_x$ (labeled the 11 system). We use Cu to substitute for Fe in Fe$_{0.98}$Te$_{0.5}$Se$_{0.5}$, and measure how both the transport properties and low-energy magnetic excitations evolve as a function of Cu concentration. With increasing Cu substitution, the system is driven towards an insulator. The low-energy ($\leq12$~meV) magnetic excitations respond to the Cu doping by showing enhanced spectral weight and longer dynamical spin-spin correlation lengths. This is in contrast to the expectation that using weakly (not) magnetic Cu to replace magnetic Fe$^{2+}$ suppresses the magnetic correlations. The behavior that we observe can be naturally explained by assuming that the main effect of Cu doping is to localize conduction states, thereby suppressing the itinerancy. As a result, either the local moments are enhanced, or the spin fluctuation rate is reduced, and the spectral weight is transferred to low energies. Either case will give rise to an enhancement of the low-energy magnetic scattering. Our results demonstrate that Cu substitution in the 11 system cannot be described by a rigid-band shift model.~\cite{PhysRevB.83.094522} 

\section{Experimental}
The single-crystal samples of Fe$_{1-y-z}$Cu$_z$Te$_{0.5}$Se$_{0.5}$ were grown by the Bridgman method.~\cite{interplaywen} We studied three samples with nominal $z=0$, 0.02, and 0.1, which we labeled as Cu0, Cu02, and Cu10. To minimize the effects of Fe interstitials, a nominal composition of $y=0.02$ was used for all three samples. From our previous X-ray and neutron powder diffraction, and inductively coupled plasma measurements on the sample compositions, the maximum deviation of the real composition from the nominal one will be less than 2\%.~\cite{xudoping11,2011arXiv1108.5968Z} The resistivity was measured with a standard four-probe method. The lattice constants at room temperature are $a=b\approx3.8$~\AA, and $c=6.1$~\AA{}, using the two-Fe unit cell. 

Neutron scattering experiments on Cu02 and Cu10 samples were carried out on the HB1 and HB3 triple-axis spectrometer at the High Flux Isotope Reactor, Oak Ridge National Laboratory. The Cu0 sample was measured on PUMA at FRM-II (Garching, Germany). On all the three spectrometers, we used the fixed final energy ($E_f$) mode with $E_f=14.7$~meV. Neutron beam collimations used on HB1 and HB3 were $48'$--$40'$--Sample--$40'$--$240'$, and $48'$--$40'$--Sample--$40'$--$120'$ respectively. All the measurements were performed in the ($HK$0) zone, with the scattering plane being defined by the [100] and [010] wave vectors, where we used reciprocal lattice units (rlu) of $(a^*, b^*,c^*)=(2\pi/a,2\pi/b,2\pi/c)$. The measured intensity ${\rm I_{meas}}$ was normalized into absolute units of $\mu_{\rm B}^2$eV$^{-1}$/Fe  by the integrated incoherent elastic scattering intensity ${\rm I_{inc}}$ from the sample, using the formula~\cite{neutron1,2013arXiv1305.5521X}
\begin{equation*}
S({\bf Q},\omega)=\frac{{\rm I_{meas}}\mu^2_{\rm B}}{4\pi{\rm {I_{inc}}}{|f(\bf Q)|^2}{\rm p^2}}\sum_jn_j\sigma_{{\rm inc}, j},
\end{equation*}
where $\mu_{\rm B}$ is the Bohr magneton, $f(\bf Q)$ is the wave-vector ($\bf Q$) dependent magnetic form factor of Fe$^{2+}$, p is a constant of 0.27$\times$10$^{-12}$~cm, $n_j$ and $\sigma_{{\rm inc}, j}$ are the molar ratio and the incoherent cross section for the element in the compound respectively.

\section{Results and Discussions}
\begin{figure}[htb]
\includegraphics[width=0.8\linewidth]{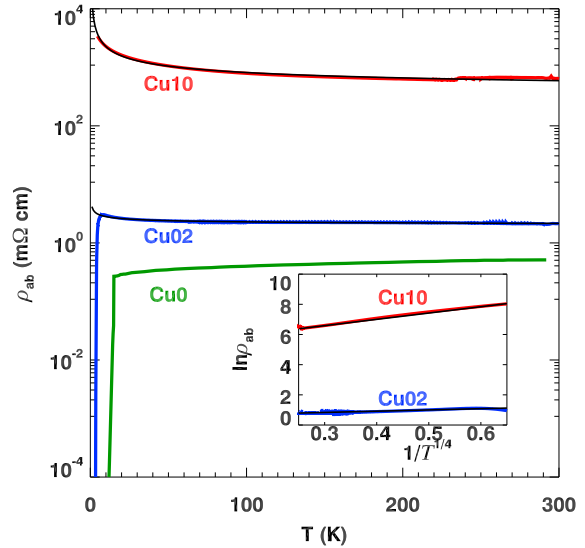}
\caption{(Color online) $a$-$b$ plane resistivity ($\rho_{\rm ab}$) vs. temperature curve in the semi-log scale for Cu0, Cu02 and Cu10 samples. The inset plots ln$\rho_{ab}$ against $1/T^{1/4}$ for Cu02 and Cu10 samples. Lines through data are fits with the three-dimensional Mott variable range hopping formula, as described in the text.~\cite{mottvariable}}
\label{rt}
\end{figure}

We first present our results by showing the $a$-$b$ plane resistivity vs. temperature ($\rho_{\rm ab}$-$T$) curves for Cu0, Cu02 and Cu10 samples in Fig.~\ref{rt}. Without Cu substitution, the sample has the highest critical temperature $T_c$ of $\sim$15~K among the Fe$_{1+y}$Te$_{1-x}$Se$_{x}$ system.~\cite{spinglass} As shown in Fig.~\ref{rt}, for Se concentrations close to 0.5, the resistivity decreases with decreasing temperature before it drops to zero at $T_c$, exhibiting a metallic behavior.~\cite{si:052504,0953-2048-24-3-035012}  The $T_c$ is suppressed rapidly by the Cu doping--with 0.02-Cu substitution, it is reduced to 8~K. In the normal state, the temperature dependence of the resistivity differs from that in the Cu-free sample. Specifically, the resistivity increases gradually with decreasing temperature, as in a narrow-band-gap semiconductor. Also, with 2\% Cu doping, the value of the resistivity has increased by almost an order of magnitude. With 10\%-Cu doping, the sample is no longer superconducting, and behaves like an insulator. The resistivity is about 4 orders of magnitude larger than that of the Cu-free sample. We fit the $a$-$b$ plane resistivity ($\rho_{ab}$) for Cu02 and Cu10 samples with the Mott variable range hopping formula $\rho_{ab}=\rho_0{\rm exp(T_0}/T^{1/(1+d)})$, where $\rho_0, T_0$ are constants, and $d$ is the dimensionality.~\cite{mottvariable} With $d=3$, the data can be fitted reasonably well, as shown in Fig.~\ref{rt} and its inset. This indicates that the Cu02 (in the normal state) and Cu10 samples behave like three-dimensional Mott insulators, similar to the behavior in Cu-doped FeSe for Cu doping larger than 4\%.~\cite{PhysRevB.82.104502,williams-2009-21}

\begin{figure}[htb]
\includegraphics[height=0.9\linewidth,angle=90]{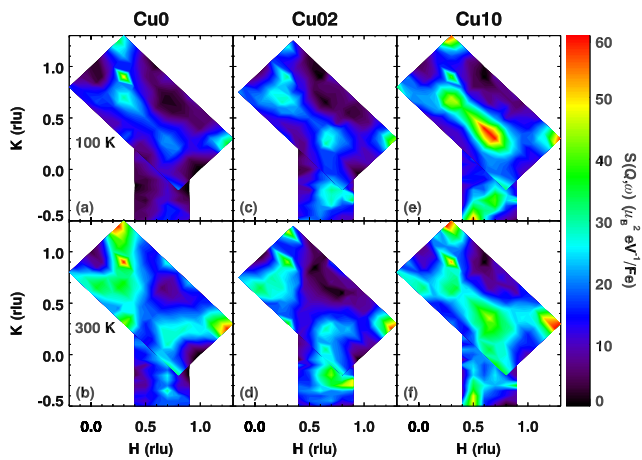}
\caption{(Color online) Contour plots of the magnetic scattering at 6~meV at 100~K (upper panels) and 300~K (bottom) for Cu0 (left column), Cu02 (middle), and Cu10 sample (right).}
\label{mesh}
\end{figure}

\begin{figure}[htb]
\includegraphics[width=0.8\linewidth]{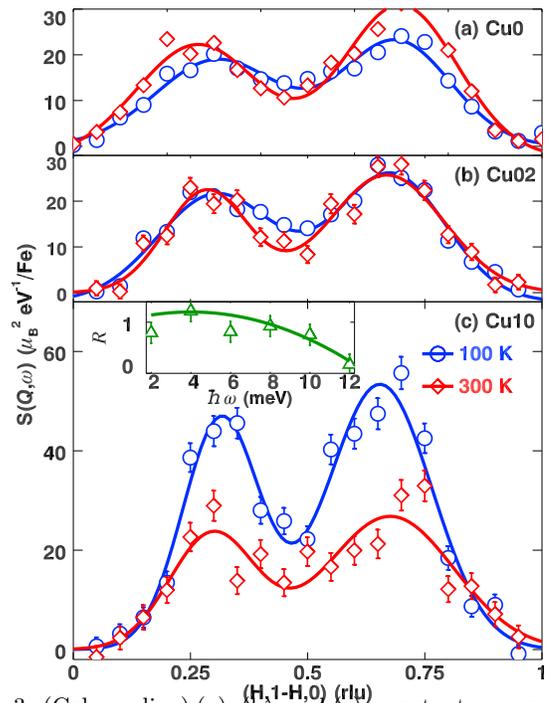}
\caption{(Color online) (a), (b), and (c), constant-energy scans of 6~meV through (0.5, 0.5) along [1$\bar{1}$0] direction at 100 and 300~K for Cu0, Cu02, and Cu10 respectively. The scan trajectories are the same as depicted in Fig.~\ref{s3100k}(a). Lines through data are fits with Gaussian functions. In the inset of (c) we plot the ratio ($R$) of enhancement on the 100-K integrated intensities (${\rm I_{100K}}$) to that of 300-K ($\rm I_{300K}$) for these scans at different energies, with $R=\rm (I_{100K}-I_{300K})/I_{300K}$. The line through data is a guide to the eyes.}
\label{tcomp}
\end{figure}

\begin{figure}[htb]
\includegraphics[height=0.8\linewidth]{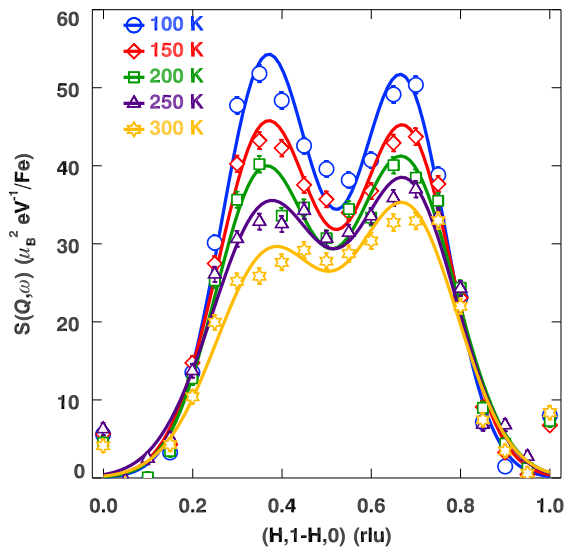}
\caption{(Color online) Linear scans through (0.5, 0.5, 0) along the [1$\bar{1}$0] direction at a constant energy of 6~meV at various temperatures for the Cu10 sample.}
\label{ppscan}
\end{figure}

Given the dramatic change of the transport properties with Cu doping, it is important to explore the corresponding response of the magnetic excitations by carrying out inelastic neutron scattering measurements. In one of our previous studies,~\cite{PhysRevLett.109.227002} we have done measurements in the low-temperature range and observed interesting connections between the occurrence of superconductivity and the shape of the magnetic excitation spectrum (see also Ref.~\onlinecite{1367-2630-14-7-073025}): for the superconducting samples, the spectra exhibit a two-vertical-line shape at high temperatures, and transform to a ``U" shape at temperature $\sim$~3$T_c$; while for the nonsuperconducting ones, the spectra remain the two-vertical-line shape in the whole temperature range. In this work, we will focus on the high temperature range with temperature $T\geq100$~K. In Fig.~\ref{mesh}, we plot contour maps for a series of scans around (0.5, 0.5) and (0.5, 0) at constant energy of 6~meV at 100 and 300~K for each of the Cu0, Cu02, and Cu10 samples. Similar as previous studies,\cite{PhysRevLett.109.227002,xudoping11,lumsden-2009,liupi0topp,PhysRevB.87.224410,1367-2630-14-7-073025} for Se content close to 50\%, there is not much spectral weight around (0.5, 0), and there is neither static magnetic order near (0.5, 0.5) nor (0.5, 0). At this temperature range, the scattering is incommensurate, with the strongest scattering occurring at wave vectors displaced from (0.5, 0.5) along the [1$\bar{1}$0] direction. From the 100-K data [Fig.~\ref{mesh}(a), (c), and (e)], it is clear that the spectral weight is greatly enhanced in the Cu10 sample compared to that of the Cu0 and Cu02 samples. For each sample, as the temperature increases from 100 to 300~K, the magnetic excitations become broader, especially along the [110] direction. However the temperature dependence of the Cu10 sample differs from that of the other two in an important way. As shown in Fig.~\ref{tcomp}, while the scattering intensity for all three samples is comparable at 300~K, only in the case of Cu10 does it grow substantially on cooling to 100~K. We have studied the energy dependence of the enhancement on the integrated intensities for the [1$\bar1$0] scan for the Cu10 sample at 100~K compared to that at 300~K, and the results are shown in the inset of Fig.~\ref{tcomp}(c). At low energies, for example, from 2 to 10~meV, the integrated intensities are almost doubled. In Fig.~\ref{ppscan}, we plot linear scans through (0.5, 0.5, 0) along the [1$\bar{1}$0] direction at a constant energy of 6~meV at temperatures ranging from 100 to 300~K, where we find that the spectral weight is enhanced gradually upon cooling. For temperatures below 100~K, the scattering intensity appears to be saturated. 

\begin{figure}[htb]
\includegraphics[width=0.9\linewidth]{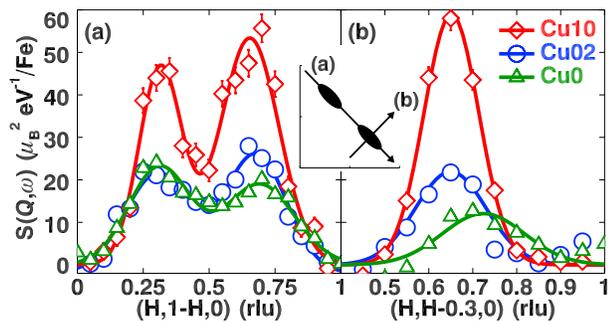}
\caption{(Color online) Linear scans at 6~meV for Cu0, Cu02 and Cu10 samples at 100~K. Lines through data are fits with Gaussian functions. The scan trajectories are shown in the inset.}
\label{s3100k}
\end{figure}

In Fig.~\ref{s3100k}(a) we plot on linear scans along [1$\bar{1}$0] direction through (0.5, 0.5, 0) at 100~K for the three samples. Comparing the scattering intensities between Cu02 and Cu0, it is apparent that there is some enhancement, but the margin is small. However, in the Cu10 sample, the intensities are almost doubled. In Fig.~\ref{s3100k}(b) we plot scans along the [110] direction through one of the two incommensurate peaks (0.7, 0.3, 0), where it is also quite clear that the peak of the Cu10 sample is much stronger. We extract the dynamical spin-spin correlation lengths at this temperature from the Gaussian fits to the [1$\bar{1}$0] scan through the peak (0.3, 0.7, 0), and (0.7, 0.3, 0), and average the correlation lengths as $\xi_T$. The correlation length extracted from the [110] scan through (0.7, 0.3, 0) is denoted as $\xi_L$. The so-obtained values for the three samples are given in Table~\ref{correlen}. For the Cu0 and Cu02 samples, the dynamical correlation lengths are basically identical. However, the Cu10 sample does appear to exhibit a longer dynamical spin-spin correlation length, especially when one looks at the scan along the [110] direction through (0.7, 0.3, 0) ($\xi_L$). We have performed similar scans at other energies from 2 to 12~meV, and the results are similar. Combining the results from Figs.~\ref{mesh} and \ref{s3100k}, and Table~\ref{correlen}, we conclude that in the Cu10 sample, the magnetic scattering is significantly strengthened. 

\begin{table}
\caption{Dynamical spin-spin correlation lengths at 100~K extracted from Gaussian fits to the [1$\bar{1}$0] scan through (0.3, 0.7, 0) and (0.7, 0.3, 0) and averaged as ($\xi_T$), and [110] scan through (0.7, 0.3, 0) ($\xi_L$)  for Cu0, Cu02, and Cu10 samples. The dynamical correlation length $\xi$ is estimated using $1/\kappa$, where $\kappa$ is the instrument resolution corrected full width at half maximum. The uncertainties of the correlation lengths are obtained from the resolution-convoluted Gaussian fits.}
\label{correlen}
\begin{tabular}{lrrr}
\hline \hline\\
     & $\qquad\qquad$ Cu0 &$\qquad\qquad$ Cu02     &$\qquad\qquad$ Cu10\\
  \hline\\
$\xi_T$ (\AA)    & 2.9$\pm0.08$ & 3.0$\pm0.07$ & 3.9$\pm0.07$\\
$\xi_L$ (\AA)    & 3.9$\pm0.22$ & 4.2$\pm0.13$ & 5.8$\pm0.16$\\  
  \hline \hline
\end{tabular}
\end{table}

Such an enhancement of the low-energy magnetic scatterings by Cu doping is intriguing. Normally, one would expect that substituting Cu for Fe would suppress the magnetic correlations. One plausible interpretation of the results is that in the Fe-based superconductors, the magnetic excitations have contributions from both localized spins and itinerant electrons,~\cite{weiguounified,johannes-2009,np_8_709,arXiv:1202.2827,PhysRevB.84.100509,0295-5075-88-1-17010} and more importantly, these two components are entangled with each other.~\cite{2011arXiv1103.5073Z} Thus, by tuning one component, one can affect the other. In our case, we infer that when the itinerancy of the system is suppressed by the addition of Cu, the local moments can be enhanced, which in turn strengthens the low-energy magnetic excitations. We note that in Fe$_{1+y}$Te, the enhancement of instantaneous moments occurred together with the transition from a more coherent to a less coherent electronic state.~\cite{2011arXiv1103.5073Z}

In the case of Cu-doped FeSe, it has also been suggested that Cu substitution introduces local moments, and when Cu doping equals 0.12, the sample exhibits a spin-glass transition.~\cite{williams-2009-21} The results are interpreted by Chadov {\it et al.} who show that Cu is in a $d^{10}$ state, and with increasing Cu doping, the system evolves from a weak-moment itinerant state to a local-moment magnet.~\cite{localm3}  Due to the Anderson localization, a metal-insulator transition can be induced. These results are consistent with our observations. 

There is another possibility, though, which is that the overall moments of the system are not enhanced, but due to the localization effects of Cu, the system tends to order magnetically. In this case, the spin fluctuation slows down, and the spectral weight is shifted to lower energies. Upon heating, the spectral weight redistributes. Carrying out time-of-flight measurements on the Cu10 sample with energy extending up to the top of the band ($>200$~meV)~\cite{lumsden-2009} will be helpful to distinguish whether the moments of the system are enhanced or not. This work is under way.

In any case, our results show clearly that the main effect of Cu doping is to introduce localization into the system, and suppress the itinerancy. This indicates limitation of applying a rigid-band shift model, where the chemical substitution is treated as charge carrier doping.~\cite{JPSJ.80.123701,PhysRevB.83.144512} There have been some X-ray  emission/absorption spectroscopy works on the Cu-doped Ba122 system, and it has been shown that the Cu 3$d$ states are located at the bottom of the valence band in a localized shell.~\cite{0953-8984-24-21-215501} This is consistent with what we see here for the Cu-doped Fe$_{1+y}$Te$_{1-x}$Se$_x$ case; this suggests that Cu may have a universal localizing effect across different Fe-based superconductor systems. 

\section{Summary}
In summary, we have found that the substitution of Fe by Cu in Fe$_{0.98}$Te$_{0.5}$Se$_{0.5}$ drives the system from a metallic to an insulating state. Concomitant with the suppression of the system's itinerancy, there is an enhancement of the low-energy magnetic excitations. Our results are consistent with the idea that Cu doping does not introduce extra carriers into the system, but rather enhances a more localized state. Such observations do not favor a simple band shift model. Given the reports that Co/Ni and Cu may act differently when substituted into Ba122,~\cite{canfield:060501,2011arXiv1112.4858B,0953-8984-24-21-215501,PhysRevLett.110.107007,PhysRevB.82.024519} it will be interesting to study the effects of Co- and Ni-doping effects on Fe$_{1+y}$Te$_{1-x}$Se$_{x}$, and specifically to compare those results with the Cu-doping effects presented here. 

\section{Acknowledgments}
We are grateful for the stimulating discussions with Wei Ku, Weiguo Yin, Tom Berlijn, Qianghua Wang, Jianxin Li, and Haihu Wen. The work at Lawrence Berkeley National Laboratory and Brookhaven National Laboratory was supported by the Office of Basic Energy Sciences, Division of Materials Science and Engineering, U.S. Department of Energy, under Contract No.\ DE-AC02-05CH11231 and DE-AC02-98CH10886 respectively. Research at Oak Ridge National Laboratory's High Flux Isotope Reactor was sponsored by the Division of Scientific User Facilities of the same Office.


\begin{thebibliography}{37}%
\makeatletter
\providecommand \@ifxundefined [1]{%
 \@ifx{#1\undefined}
}%
\providecommand \@ifnum [1]{%
 \ifnum #1\expandafter \@firstoftwo
 \else \expandafter \@secondoftwo
 \fi
}%
\providecommand \@ifx [1]{%
 \ifx #1\expandafter \@firstoftwo
 \else \expandafter \@secondoftwo
 \fi
}%
\providecommand \natexlab [1]{#1}%
\providecommand \enquote  [1]{``#1''}%
\providecommand \bibnamefont  [1]{#1}%
\providecommand \bibfnamefont [1]{#1}%
\providecommand \citenamefont [1]{#1}%
\providecommand \href@noop [0]{\@secondoftwo}%
\providecommand \href [0]{\begingroup \@sanitize@url \@href}%
\providecommand \@href[1]{\@@startlink{#1}\@@href}%
\providecommand \@@href[1]{\endgroup#1\@@endlink}%
\providecommand \@sanitize@url [0]{\catcode `\\12\catcode `\$12\catcode
  `\&12\catcode `\#12\catcode `\^12\catcode `\_12\catcode `\%12\relax}%
\providecommand \@@startlink[1]{}%
\providecommand \@@endlink[0]{}%
\providecommand \url  [0]{\begingroup\@sanitize@url \@url }%
\providecommand \@url [1]{\endgroup\@href {#1}{\urlprefix }}%
\providecommand \urlprefix  [0]{URL }%
\providecommand \Eprint [0]{\href }%
\providecommand \doibase [0]{http://dx.doi.org/}%
\providecommand \selectlanguage [0]{\@gobble}%
\providecommand \bibinfo  [0]{\@secondoftwo}%
\providecommand \bibfield  [0]{\@secondoftwo}%
\providecommand \translation [1]{[#1]}%
\providecommand \BibitemOpen [0]{}%
\providecommand \bibitemStop [0]{}%
\providecommand \bibitemNoStop [0]{.\EOS\space}%
\providecommand \EOS [0]{\spacefactor3000\relax}%
\providecommand \BibitemShut  [1]{\csname bibitem#1\endcsname}%
\let\auto@bib@innerbib\@empty
\bibitem [{\citenamefont {Canfield}\ \emph {et~al.}(2009)\citenamefont
  {Canfield}, \citenamefont {Bud'ko}, \citenamefont {Ni}, \citenamefont {Yan},\
  and\ \citenamefont {Kracher}}]{canfield:060501}%
  \BibitemOpen
  \bibfield  {author} {\bibinfo {author} {\bibfnamefont {P.~C.}\ \bibnamefont
  {Canfield}}, \bibinfo {author} {\bibfnamefont {S.~L.}\ \bibnamefont
  {Bud'ko}}, \bibinfo {author} {\bibfnamefont {N.}~\bibnamefont {Ni}}, \bibinfo
  {author} {\bibfnamefont {J.~Q.}\ \bibnamefont {Yan}}, \ and\ \bibinfo
  {author} {\bibfnamefont {A.}~\bibnamefont {Kracher}},\ }\href@noop {}
  {\bibfield  {journal} {\bibinfo  {journal} {Phys. Rev. B}\ }\textbf {\bibinfo
  {volume} {80}},\ \bibinfo {pages} {060501 (R)} (\bibinfo {year}
  {2009})}\BibitemShut {NoStop}%
\bibitem [{\citenamefont {Ideta}\ \emph {et~al.}(2013)\citenamefont {Ideta},
  \citenamefont {Yoshida}, \citenamefont {Nishi}, \citenamefont {Fujimori},
  \citenamefont {Kotani}, \citenamefont {Ono}, \citenamefont {Nakashima},
  \citenamefont {Yamaichi}, \citenamefont {Sasagawa}, \citenamefont {Nakajima},
  \citenamefont {Kihou}, \citenamefont {Tomioka}, \citenamefont {Lee},
  \citenamefont {Iyo}, \citenamefont {Eisaki}, \citenamefont {Ito},
  \citenamefont {Uchida},\ and\ \citenamefont
  {Arita}}]{PhysRevLett.110.107007}%
  \BibitemOpen
  \bibfield  {author} {\bibinfo {author} {\bibfnamefont {S.}~\bibnamefont
  {Ideta}}, \bibinfo {author} {\bibfnamefont {T.}~\bibnamefont {Yoshida}},
  \bibinfo {author} {\bibfnamefont {I.}~\bibnamefont {Nishi}}, \bibinfo
  {author} {\bibfnamefont {A.}~\bibnamefont {Fujimori}}, \bibinfo {author}
  {\bibfnamefont {Y.}~\bibnamefont {Kotani}}, \bibinfo {author} {\bibfnamefont
  {K.}~\bibnamefont {Ono}}, \bibinfo {author} {\bibfnamefont {Y.}~\bibnamefont
  {Nakashima}}, \bibinfo {author} {\bibfnamefont {S.}~\bibnamefont {Yamaichi}},
  \bibinfo {author} {\bibfnamefont {T.}~\bibnamefont {Sasagawa}}, \bibinfo
  {author} {\bibfnamefont {M.}~\bibnamefont {Nakajima}}, \bibinfo {author}
  {\bibfnamefont {K.}~\bibnamefont {Kihou}}, \bibinfo {author} {\bibfnamefont
  {Y.}~\bibnamefont {Tomioka}}, \bibinfo {author} {\bibfnamefont {C.~H.}\
  \bibnamefont {Lee}}, \bibinfo {author} {\bibfnamefont {A.}~\bibnamefont
  {Iyo}}, \bibinfo {author} {\bibfnamefont {H.}~\bibnamefont {Eisaki}},
  \bibinfo {author} {\bibfnamefont {T.}~\bibnamefont {Ito}}, \bibinfo {author}
  {\bibfnamefont {S.}~\bibnamefont {Uchida}}, \ and\ \bibinfo {author}
  {\bibfnamefont {R.}~\bibnamefont {Arita}},\ }\href {\doibase
  10.1103/PhysRevLett.110.107007} {\bibfield  {journal} {\bibinfo  {journal}
  {Phys. Rev. Lett.}\ }\textbf {\bibinfo {volume} {110}},\ \bibinfo {pages}
  {107007} (\bibinfo {year} {2013})}\BibitemShut {NoStop}%
\bibitem [{\citenamefont {Levy}\ \emph {et~al.}(2012)\citenamefont {Levy},
  \citenamefont {Sutarto}, \citenamefont {Chevrier}, \citenamefont {Regier},
  \citenamefont {Blyth}, \citenamefont {Geck}, \citenamefont {Wurmehl},
  \citenamefont {Harnagea}, \citenamefont {Wadati}, \citenamefont {Mizokawa},
  \citenamefont {Elfimov}, \citenamefont {Damascelli},\ and\ \citenamefont
  {Sawatzky}}]{PhysRevLett.109.077001}%
  \BibitemOpen
  \bibfield  {author} {\bibinfo {author} {\bibfnamefont {G.}~\bibnamefont
  {Levy}}, \bibinfo {author} {\bibfnamefont {R.}~\bibnamefont {Sutarto}},
  \bibinfo {author} {\bibfnamefont {D.}~\bibnamefont {Chevrier}}, \bibinfo
  {author} {\bibfnamefont {T.}~\bibnamefont {Regier}}, \bibinfo {author}
  {\bibfnamefont {R.}~\bibnamefont {Blyth}}, \bibinfo {author} {\bibfnamefont
  {J.}~\bibnamefont {Geck}}, \bibinfo {author} {\bibfnamefont {S.}~\bibnamefont
  {Wurmehl}}, \bibinfo {author} {\bibfnamefont {L.}~\bibnamefont {Harnagea}},
  \bibinfo {author} {\bibfnamefont {H.}~\bibnamefont {Wadati}}, \bibinfo
  {author} {\bibfnamefont {T.}~\bibnamefont {Mizokawa}}, \bibinfo {author}
  {\bibfnamefont {I.~S.}\ \bibnamefont {Elfimov}}, \bibinfo {author}
  {\bibfnamefont {A.}~\bibnamefont {Damascelli}}, \ and\ \bibinfo {author}
  {\bibfnamefont {G.~A.}\ \bibnamefont {Sawatzky}},\ }\href {\doibase
  10.1103/PhysRevLett.109.077001} {\bibfield  {journal} {\bibinfo  {journal}
  {Phys. Rev. Lett.}\ }\textbf {\bibinfo {volume} {109}},\ \bibinfo {pages}
  {077001} (\bibinfo {year} {2012})}\BibitemShut {NoStop}%
\bibitem [{\citenamefont {Liu}\ \emph {et~al.}(2011{\natexlab{a}})\citenamefont
  {Liu}, \citenamefont {Palczewski}, \citenamefont {Dhaka}, \citenamefont
  {Kondo}, \citenamefont {Fernandes}, \citenamefont {Mun}, \citenamefont
  {Hodovanets}, \citenamefont {Thaler}, \citenamefont {Schmalian},
  \citenamefont {Bud'ko}, \citenamefont {Canfield},\ and\ \citenamefont
  {Kaminski}}]{PhysRevB.84.020509}%
  \BibitemOpen
  \bibfield  {author} {\bibinfo {author} {\bibfnamefont {C.}~\bibnamefont
  {Liu}}, \bibinfo {author} {\bibfnamefont {A.~D.}\ \bibnamefont {Palczewski}},
  \bibinfo {author} {\bibfnamefont {R.~S.}\ \bibnamefont {Dhaka}}, \bibinfo
  {author} {\bibfnamefont {T.}~\bibnamefont {Kondo}}, \bibinfo {author}
  {\bibfnamefont {R.~M.}\ \bibnamefont {Fernandes}}, \bibinfo {author}
  {\bibfnamefont {E.~D.}\ \bibnamefont {Mun}}, \bibinfo {author} {\bibfnamefont
  {H.}~\bibnamefont {Hodovanets}}, \bibinfo {author} {\bibfnamefont {A.~N.}\
  \bibnamefont {Thaler}}, \bibinfo {author} {\bibfnamefont {J.}~\bibnamefont
  {Schmalian}}, \bibinfo {author} {\bibfnamefont {S.~L.}\ \bibnamefont
  {Bud'ko}}, \bibinfo {author} {\bibfnamefont {P.~C.}\ \bibnamefont
  {Canfield}}, \ and\ \bibinfo {author} {\bibfnamefont {A.}~\bibnamefont
  {Kaminski}},\ }\href {\doibase 10.1103/PhysRevB.84.020509} {\bibfield
  {journal} {\bibinfo  {journal} {Phys. Rev. B}\ }\textbf {\bibinfo {volume}
  {84}},\ \bibinfo {pages} {020509} (\bibinfo {year}
  {2011}{\natexlab{a}})}\BibitemShut {NoStop}%
\bibitem [{\citenamefont {Ni}\ \emph {et~al.}(2010)\citenamefont {Ni},
  \citenamefont {Thaler}, \citenamefont {Yan}, \citenamefont {Kracher},
  \citenamefont {Colombier}, \citenamefont {Bud'ko}, \citenamefont {Canfield},\
  and\ \citenamefont {Hannahs}}]{PhysRevB.82.024519}%
  \BibitemOpen
  \bibfield  {author} {\bibinfo {author} {\bibfnamefont {N.}~\bibnamefont
  {Ni}}, \bibinfo {author} {\bibfnamefont {A.}~\bibnamefont {Thaler}}, \bibinfo
  {author} {\bibfnamefont {J.~Q.}\ \bibnamefont {Yan}}, \bibinfo {author}
  {\bibfnamefont {A.}~\bibnamefont {Kracher}}, \bibinfo {author} {\bibfnamefont
  {E.}~\bibnamefont {Colombier}}, \bibinfo {author} {\bibfnamefont {S.~L.}\
  \bibnamefont {Bud'ko}}, \bibinfo {author} {\bibfnamefont {P.~C.}\
  \bibnamefont {Canfield}}, \ and\ \bibinfo {author} {\bibfnamefont {S.~T.}\
  \bibnamefont {Hannahs}},\ }\href {\doibase 10.1103/PhysRevB.82.024519}
  {\bibfield  {journal} {\bibinfo  {journal} {Phys. Rev. B}\ }\textbf {\bibinfo
  {volume} {82}},\ \bibinfo {pages} {024519} (\bibinfo {year}
  {2010})}\BibitemShut {NoStop}%
\bibitem [{\citenamefont {Neupane}\ \emph {et~al.}(2011)\citenamefont
  {Neupane}, \citenamefont {Richard}, \citenamefont {Xu}, \citenamefont
  {Nakayama}, \citenamefont {Sato}, \citenamefont {Takahashi}, \citenamefont
  {Federov}, \citenamefont {Xu}, \citenamefont {Dai}, \citenamefont {Fang},
  \citenamefont {Wang}, \citenamefont {Chen}, \citenamefont {Wang},
  \citenamefont {Wen},\ and\ \citenamefont {Ding}}]{PhysRevB.83.094522}%
  \BibitemOpen
  \bibfield  {author} {\bibinfo {author} {\bibfnamefont {M.}~\bibnamefont
  {Neupane}}, \bibinfo {author} {\bibfnamefont {P.}~\bibnamefont {Richard}},
  \bibinfo {author} {\bibfnamefont {Y.-M.}\ \bibnamefont {Xu}}, \bibinfo
  {author} {\bibfnamefont {K.}~\bibnamefont {Nakayama}}, \bibinfo {author}
  {\bibfnamefont {T.}~\bibnamefont {Sato}}, \bibinfo {author} {\bibfnamefont
  {T.}~\bibnamefont {Takahashi}}, \bibinfo {author} {\bibfnamefont {A.~V.}\
  \bibnamefont {Federov}}, \bibinfo {author} {\bibfnamefont {G.}~\bibnamefont
  {Xu}}, \bibinfo {author} {\bibfnamefont {X.}~\bibnamefont {Dai}}, \bibinfo
  {author} {\bibfnamefont {Z.}~\bibnamefont {Fang}}, \bibinfo {author}
  {\bibfnamefont {Z.}~\bibnamefont {Wang}}, \bibinfo {author} {\bibfnamefont
  {G.-F.}\ \bibnamefont {Chen}}, \bibinfo {author} {\bibfnamefont {N.-L.}\
  \bibnamefont {Wang}}, \bibinfo {author} {\bibfnamefont {H.-H.}\ \bibnamefont
  {Wen}}, \ and\ \bibinfo {author} {\bibfnamefont {H.}~\bibnamefont {Ding}},\
  }\href {\doibase 10.1103/PhysRevB.83.094522} {\bibfield  {journal} {\bibinfo
  {journal} {Phys. Rev. B}\ }\textbf {\bibinfo {volume} {83}},\ \bibinfo
  {pages} {094522} (\bibinfo {year} {2011})}\BibitemShut {NoStop}%
\bibitem [{\citenamefont {Nakamura}\ \emph {et~al.}(2011)\citenamefont
  {Nakamura}, \citenamefont {Arita},\ and\ \citenamefont
  {Ikeda}}]{PhysRevB.83.144512}%
  \BibitemOpen
  \bibfield  {author} {\bibinfo {author} {\bibfnamefont {K.}~\bibnamefont
  {Nakamura}}, \bibinfo {author} {\bibfnamefont {R.}~\bibnamefont {Arita}}, \
  and\ \bibinfo {author} {\bibfnamefont {H.}~\bibnamefont {Ikeda}},\ }\href
  {\doibase 10.1103/PhysRevB.83.144512} {\bibfield  {journal} {\bibinfo
  {journal} {Phys. Rev. B}\ }\textbf {\bibinfo {volume} {83}},\ \bibinfo
  {pages} {144512} (\bibinfo {year} {2011})}\BibitemShut {NoStop}%
\bibitem [{\citenamefont {Konbu}\ \emph {et~al.}(2011)\citenamefont {Konbu},
  \citenamefont {Nakamura}, \citenamefont {Ikeda},\ and\ \citenamefont
  {Arita}}]{JPSJ.80.123701}%
  \BibitemOpen
  \bibfield  {author} {\bibinfo {author} {\bibfnamefont {S.}~\bibnamefont
  {Konbu}}, \bibinfo {author} {\bibfnamefont {K.}~\bibnamefont {Nakamura}},
  \bibinfo {author} {\bibfnamefont {H.}~\bibnamefont {Ikeda}}, \ and\ \bibinfo
  {author} {\bibfnamefont {R.}~\bibnamefont {Arita}},\ }\href {\doibase
  10.1143/JPSJ.80.123701} {\bibfield  {journal} {\bibinfo  {journal} {Journal
  of the Physical Society of Japan}\ }\textbf {\bibinfo {volume} {80}},\
  \bibinfo {pages} {123701} (\bibinfo {year} {2011})}\BibitemShut {NoStop}%
\bibitem [{\citenamefont {Wadati}\ \emph {et~al.}(2010)\citenamefont {Wadati},
  \citenamefont {Elfimov},\ and\ \citenamefont
  {Sawatzky}}]{PhysRevLett.105.157004}%
  \BibitemOpen
  \bibfield  {author} {\bibinfo {author} {\bibfnamefont {H.}~\bibnamefont
  {Wadati}}, \bibinfo {author} {\bibfnamefont {I.}~\bibnamefont {Elfimov}}, \
  and\ \bibinfo {author} {\bibfnamefont {G.~A.}\ \bibnamefont {Sawatzky}},\
  }\href {\doibase 10.1103/PhysRevLett.105.157004} {\bibfield  {journal}
  {\bibinfo  {journal} {Phys. Rev. Lett.}\ }\textbf {\bibinfo {volume} {105}},\
  \bibinfo {pages} {157004} (\bibinfo {year} {2010})}\BibitemShut {NoStop}%
\bibitem [{\citenamefont {Berlijn}\ \emph {et~al.}(2012)\citenamefont
  {Berlijn}, \citenamefont {Lin}, \citenamefont {Garber},\ and\ \citenamefont
  {Ku}}]{2011arXiv1112.4858B}%
  \BibitemOpen
  \bibfield  {author} {\bibinfo {author} {\bibfnamefont {T.}~\bibnamefont
  {Berlijn}}, \bibinfo {author} {\bibfnamefont {C.-H.}\ \bibnamefont {Lin}},
  \bibinfo {author} {\bibfnamefont {W.}~\bibnamefont {Garber}}, \ and\ \bibinfo
  {author} {\bibfnamefont {W.}~\bibnamefont {Ku}},\ }\href {\doibase
  10.1103/PhysRevLett.108.207003} {\bibfield  {journal} {\bibinfo  {journal}
  {Phys. Rev. Lett.}\ }\textbf {\bibinfo {volume} {108}},\ \bibinfo {pages}
  {207003} (\bibinfo {year} {2012})}\BibitemShut {NoStop}%
\bibitem [{\citenamefont {Bittar}\ \emph {et~al.}(2011)\citenamefont {Bittar},
  \citenamefont {Adriano}, \citenamefont {Garitezi}, \citenamefont {Rosa},
  \citenamefont {Mendon\ifmmode \mbox{\c{c}}\else~\c{c}\fi{}a Ferreira},
  \citenamefont {Garcia}, \citenamefont {Azevedo}, \citenamefont {Pagliuso},\
  and\ \citenamefont {Granado}}]{2011arXiv1107.0962B}%
  \BibitemOpen
  \bibfield  {author} {\bibinfo {author} {\bibfnamefont {E.~M.}\ \bibnamefont
  {Bittar}}, \bibinfo {author} {\bibfnamefont {C.}~\bibnamefont {Adriano}},
  \bibinfo {author} {\bibfnamefont {T.~M.}\ \bibnamefont {Garitezi}}, \bibinfo
  {author} {\bibfnamefont {P.~F.~S.}\ \bibnamefont {Rosa}}, \bibinfo {author}
  {\bibfnamefont {L.}~\bibnamefont {Mendon\ifmmode
  \mbox{\c{c}}\else~\c{c}\fi{}a Ferreira}}, \bibinfo {author} {\bibfnamefont
  {F.}~\bibnamefont {Garcia}}, \bibinfo {author} {\bibfnamefont {G.~d.~M.}\
  \bibnamefont {Azevedo}}, \bibinfo {author} {\bibfnamefont {P.~G.}\
  \bibnamefont {Pagliuso}}, \ and\ \bibinfo {author} {\bibfnamefont
  {E.}~\bibnamefont {Granado}},\ }\href {\doibase
  10.1103/PhysRevLett.107.267402} {\bibfield  {journal} {\bibinfo  {journal}
  {Phys. Rev. Lett.}\ }\textbf {\bibinfo {volume} {107}},\ \bibinfo {pages}
  {267402} (\bibinfo {year} {2011})}\BibitemShut {NoStop}%
\bibitem [{\citenamefont {Kuo}\ \emph {et~al.}(2011)\citenamefont {Kuo},
  \citenamefont {Chu}, \citenamefont {Riggs}, \citenamefont {Yu}, \citenamefont
  {McMahon}, \citenamefont {De~Greve}, \citenamefont {Yamamoto}, \citenamefont
  {Analytis},\ and\ \citenamefont {Fisher}}]{PhysRevB.84.054540}%
  \BibitemOpen
  \bibfield  {author} {\bibinfo {author} {\bibfnamefont {H.-H.}\ \bibnamefont
  {Kuo}}, \bibinfo {author} {\bibfnamefont {J.-H.}\ \bibnamefont {Chu}},
  \bibinfo {author} {\bibfnamefont {S.~C.}\ \bibnamefont {Riggs}}, \bibinfo
  {author} {\bibfnamefont {L.}~\bibnamefont {Yu}}, \bibinfo {author}
  {\bibfnamefont {P.~L.}\ \bibnamefont {McMahon}}, \bibinfo {author}
  {\bibfnamefont {K.}~\bibnamefont {De~Greve}}, \bibinfo {author}
  {\bibfnamefont {Y.}~\bibnamefont {Yamamoto}}, \bibinfo {author}
  {\bibfnamefont {J.~G.}\ \bibnamefont {Analytis}}, \ and\ \bibinfo {author}
  {\bibfnamefont {I.~R.}\ \bibnamefont {Fisher}},\ }\href {\doibase
  10.1103/PhysRevB.84.054540} {\bibfield  {journal} {\bibinfo  {journal} {Phys.
  Rev. B}\ }\textbf {\bibinfo {volume} {84}},\ \bibinfo {pages} {054540}
  (\bibinfo {year} {2011})}\BibitemShut {NoStop}%
\bibitem [{\citenamefont {Wen}\ \emph {et~al.}(2011)\citenamefont {Wen},
  \citenamefont {Xu}, \citenamefont {Gu}, \citenamefont {Tranquada},\ and\
  \citenamefont {Birgeneau}}]{interplaywen}%
  \BibitemOpen
  \bibfield  {author} {\bibinfo {author} {\bibfnamefont {J.}~\bibnamefont
  {Wen}}, \bibinfo {author} {\bibfnamefont {G.}~\bibnamefont {Xu}}, \bibinfo
  {author} {\bibfnamefont {G.}~\bibnamefont {Gu}}, \bibinfo {author}
  {\bibfnamefont {J.~M.}\ \bibnamefont {Tranquada}}, \ and\ \bibinfo {author}
  {\bibfnamefont {R.~J.}\ \bibnamefont {Birgeneau}},\ }\href@noop {} {\bibfield
   {journal} {\bibinfo  {journal} {Rep. Pro. Phys.}\ }\textbf {\bibinfo
  {volume} {74}},\ \bibinfo {pages} {124503} (\bibinfo {year}
  {2011})}\BibitemShut {NoStop}%
\bibitem [{\citenamefont {Xu}\ \emph {et~al.}(2010)\citenamefont {Xu},
  \citenamefont {Wen}, \citenamefont {Xu}, \citenamefont {Jie}, \citenamefont
  {Lin}, \citenamefont {Li}, \citenamefont {Chi}, \citenamefont {Singh},
  \citenamefont {Gu},\ and\ \citenamefont {Tranquada}}]{xudoping11}%
  \BibitemOpen
  \bibfield  {author} {\bibinfo {author} {\bibfnamefont {Z.}~\bibnamefont
  {Xu}}, \bibinfo {author} {\bibfnamefont {J.}~\bibnamefont {Wen}}, \bibinfo
  {author} {\bibfnamefont {G.}~\bibnamefont {Xu}}, \bibinfo {author}
  {\bibfnamefont {Q.}~\bibnamefont {Jie}}, \bibinfo {author} {\bibfnamefont
  {Z.}~\bibnamefont {Lin}}, \bibinfo {author} {\bibfnamefont {Q.}~\bibnamefont
  {Li}}, \bibinfo {author} {\bibfnamefont {S.}~\bibnamefont {Chi}}, \bibinfo
  {author} {\bibfnamefont {D.~K.}\ \bibnamefont {Singh}}, \bibinfo {author}
  {\bibfnamefont {G.}~\bibnamefont {Gu}}, \ and\ \bibinfo {author}
  {\bibfnamefont {J.~M.}\ \bibnamefont {Tranquada}},\ }\href@noop {} {\bibfield
   {journal} {\bibinfo  {journal} {Phys. Rev. B}\ }\textbf {\bibinfo {volume}
  {82}},\ \bibinfo {pages} {104525} (\bibinfo {year} {2010})}\BibitemShut
  {NoStop}%
\bibitem [{\citenamefont {Zaliznyak}\ \emph {et~al.}(2012)\citenamefont
  {Zaliznyak}, \citenamefont {Xu}, \citenamefont {Wen}, \citenamefont
  {Tranquada}, \citenamefont {Gu}, \citenamefont {Solovyov}, \citenamefont
  {Glazkov}, \citenamefont {Zheludev}, \citenamefont {Garlea},\ and\
  \citenamefont {Stone}}]{2011arXiv1108.5968Z}%
  \BibitemOpen
  \bibfield  {author} {\bibinfo {author} {\bibfnamefont {I.~A.}\ \bibnamefont
  {Zaliznyak}}, \bibinfo {author} {\bibfnamefont {Z.~J.}\ \bibnamefont {Xu}},
  \bibinfo {author} {\bibfnamefont {J.~S.}\ \bibnamefont {Wen}}, \bibinfo
  {author} {\bibfnamefont {J.~M.}\ \bibnamefont {Tranquada}}, \bibinfo {author}
  {\bibfnamefont {G.~D.}\ \bibnamefont {Gu}}, \bibinfo {author} {\bibfnamefont
  {V.}~\bibnamefont {Solovyov}}, \bibinfo {author} {\bibfnamefont {V.~N.}\
  \bibnamefont {Glazkov}}, \bibinfo {author} {\bibfnamefont {A.~I.}\
  \bibnamefont {Zheludev}}, \bibinfo {author} {\bibfnamefont {V.~O.}\
  \bibnamefont {Garlea}}, \ and\ \bibinfo {author} {\bibfnamefont {M.~B.}\
  \bibnamefont {Stone}},\ }\href {\doibase 10.1103/PhysRevB.85.085105}
  {\bibfield  {journal} {\bibinfo  {journal} {Phys. Rev. B}\ }\textbf {\bibinfo
  {volume} {85}},\ \bibinfo {pages} {085105} (\bibinfo {year}
  {2012})}\BibitemShut {NoStop}%
\bibitem [{\citenamefont {Shirane}\ \emph {et~al.}(2002)\citenamefont
  {Shirane}, \citenamefont {Shapiro},\ and\ \citenamefont
  {Tranquada}}]{neutron1}%
  \BibitemOpen
  \bibfield  {author} {\bibinfo {author} {\bibfnamefont {G.}~\bibnamefont
  {Shirane}}, \bibinfo {author} {\bibfnamefont {S.~M.}\ \bibnamefont
  {Shapiro}}, \ and\ \bibinfo {author} {\bibfnamefont {J.~M.}\ \bibnamefont
  {Tranquada}},\ }\href@noop {} {\emph {\bibinfo {title} {Neutron Scattering
  with a Triple-Axis Spectrometer: Basic Techniques}}}\ (\bibinfo  {publisher}
  {Cambridge University Press},\ \bibinfo {address} {Cambridge},\ \bibinfo
  {year} {2002})\BibitemShut {NoStop}%
\bibitem [{\citenamefont {{Xu}}\ \emph {et~al.}(2013)\citenamefont {{Xu}},
  \citenamefont {{Xu}},\ and\ \citenamefont
  {{Tranquada}}}]{2013arXiv1305.5521X}%
  \BibitemOpen
  \bibfield  {author} {\bibinfo {author} {\bibfnamefont {G.}~\bibnamefont
  {{Xu}}}, \bibinfo {author} {\bibfnamefont {Z.}~\bibnamefont {{Xu}}}, \ and\
  \bibinfo {author} {\bibfnamefont {J.~M.}\ \bibnamefont {{Tranquada}}},\
  }\href@noop {} {\bibfield  {journal} {\bibinfo  {journal} {arXiv:1305.5521}\
  } (\bibinfo {year} {2013})}\BibitemShut {NoStop}%
\bibitem [{\citenamefont {Katayama}\ \emph {et~al.}(2010)\citenamefont
  {Katayama}, \citenamefont {Ji}, \citenamefont {Louca}, \citenamefont {Lee},
  \citenamefont {Fujita}, \citenamefont {Sato}, \citenamefont {Wen},
  \citenamefont {Xu}, \citenamefont {Gu}, \citenamefont {Xu}, \citenamefont
  {Lin}, \citenamefont {Enoki}, \citenamefont {Chang}, \citenamefont {Yamada},\
  and\ \citenamefont {Tranquada}}]{spinglass}%
  \BibitemOpen
  \bibfield  {author} {\bibinfo {author} {\bibfnamefont {N.}~\bibnamefont
  {Katayama}}, \bibinfo {author} {\bibfnamefont {S.}~\bibnamefont {Ji}},
  \bibinfo {author} {\bibfnamefont {D.}~\bibnamefont {Louca}}, \bibinfo
  {author} {\bibfnamefont {S.-H.}\ \bibnamefont {Lee}}, \bibinfo {author}
  {\bibfnamefont {M.}~\bibnamefont {Fujita}}, \bibinfo {author} {\bibfnamefont
  {T.~J.}\ \bibnamefont {Sato}}, \bibinfo {author} {\bibfnamefont {J.~S.}\
  \bibnamefont {Wen}}, \bibinfo {author} {\bibfnamefont {Z.~J.}\ \bibnamefont
  {Xu}}, \bibinfo {author} {\bibfnamefont {G.~D.}\ \bibnamefont {Gu}}, \bibinfo
  {author} {\bibfnamefont {G.}~\bibnamefont {Xu}}, \bibinfo {author}
  {\bibfnamefont {Z.~W.}\ \bibnamefont {Lin}}, \bibinfo {author} {\bibfnamefont
  {M.}~\bibnamefont {Enoki}}, \bibinfo {author} {\bibfnamefont
  {S.}~\bibnamefont {Chang}}, \bibinfo {author} {\bibfnamefont
  {K.}~\bibnamefont {Yamada}}, \ and\ \bibinfo {author} {\bibfnamefont {J.~M.}\
  \bibnamefont {Tranquada}},\ }\href@noop {} {\bibfield  {journal} {\bibinfo
  {journal} {J. Phys. Soc. Jpn.}\ }\textbf {\bibinfo {volume} {79}},\ \bibinfo
  {pages} {113702} (\bibinfo {year} {2010})}\BibitemShut {NoStop}%
\bibitem [{\citenamefont {Si}\ \emph {et~al.}(2009)\citenamefont {Si},
  \citenamefont {Lin}, \citenamefont {Jie}, \citenamefont {Yin}, \citenamefont
  {Zhou}, \citenamefont {Gu}, \citenamefont {Johnson},\ and\ \citenamefont
  {Li}}]{si:052504}%
  \BibitemOpen
  \bibfield  {author} {\bibinfo {author} {\bibfnamefont {W.}~\bibnamefont
  {Si}}, \bibinfo {author} {\bibfnamefont {Z.-W.}\ \bibnamefont {Lin}},
  \bibinfo {author} {\bibfnamefont {Q.}~\bibnamefont {Jie}}, \bibinfo {author}
  {\bibfnamefont {W.-G.}\ \bibnamefont {Yin}}, \bibinfo {author} {\bibfnamefont
  {J.}~\bibnamefont {Zhou}}, \bibinfo {author} {\bibfnamefont {G.}~\bibnamefont
  {Gu}}, \bibinfo {author} {\bibfnamefont {P.~D.}\ \bibnamefont {Johnson}}, \
  and\ \bibinfo {author} {\bibfnamefont {Q.}~\bibnamefont {Li}},\ }\href@noop
  {} {\bibfield  {journal} {\bibinfo  {journal} {Appl. Phys. Lett.}\ }\textbf
  {\bibinfo {volume} {95}},\ \bibinfo {pages} {052504} (\bibinfo {year}
  {2009})}\BibitemShut {NoStop}%
\bibitem [{\citenamefont {Liu}\ \emph {et~al.}(2011{\natexlab{b}})\citenamefont
  {Liu}, \citenamefont {Kremer},\ and\ \citenamefont
  {Lin}}]{0953-2048-24-3-035012}%
  \BibitemOpen
  \bibfield  {author} {\bibinfo {author} {\bibfnamefont {Y.}~\bibnamefont
  {Liu}}, \bibinfo {author} {\bibfnamefont {R.~K.}\ \bibnamefont {Kremer}}, \
  and\ \bibinfo {author} {\bibfnamefont {C.~T.}\ \bibnamefont {Lin}},\ }\href
  {http://stacks.iop.org/0953-2048/24/i=3/a=035012} {\bibfield  {journal}
  {\bibinfo  {journal} {Supercond. Sci. Tech.}\ }\textbf {\bibinfo {volume}
  {24}},\ \bibinfo {pages} {035012} (\bibinfo {year}
  {2011}{\natexlab{b}})}\BibitemShut {NoStop}%
\bibitem [{\citenamefont {Mott}(1969)}]{mottvariable}%
  \BibitemOpen
  \bibfield  {author} {\bibinfo {author} {\bibfnamefont {N.~F.}\ \bibnamefont
  {Mott}},\ }\href {\doibase 10.1080/14786436908216338} {\bibfield  {journal}
  {\bibinfo  {journal} {Phil. Mag.}\ }\textbf {\bibinfo {volume} {19}},\
  \bibinfo {pages} {835} (\bibinfo {year} {1969})}\BibitemShut {NoStop}%
\bibitem [{\citenamefont {Huang}\ \emph {et~al.}(2010)\citenamefont {Huang},
  \citenamefont {Chen}, \citenamefont {Yeh}, \citenamefont {Ke}, \citenamefont
  {Chen}, \citenamefont {Huang}, \citenamefont {Hsu}, \citenamefont {Wu},
  \citenamefont {Wu}, \citenamefont {Avdeev},\ and\ \citenamefont
  {Studer}}]{PhysRevB.82.104502}%
  \BibitemOpen
  \bibfield  {author} {\bibinfo {author} {\bibfnamefont {T.-W.}\ \bibnamefont
  {Huang}}, \bibinfo {author} {\bibfnamefont {T.-K.}\ \bibnamefont {Chen}},
  \bibinfo {author} {\bibfnamefont {K.-W.}\ \bibnamefont {Yeh}}, \bibinfo
  {author} {\bibfnamefont {C.-T.}\ \bibnamefont {Ke}}, \bibinfo {author}
  {\bibfnamefont {C.~L.}\ \bibnamefont {Chen}}, \bibinfo {author}
  {\bibfnamefont {Y.-L.}\ \bibnamefont {Huang}}, \bibinfo {author}
  {\bibfnamefont {F.-C.}\ \bibnamefont {Hsu}}, \bibinfo {author} {\bibfnamefont
  {M.-K.}\ \bibnamefont {Wu}}, \bibinfo {author} {\bibfnamefont {P.~M.}\
  \bibnamefont {Wu}}, \bibinfo {author} {\bibfnamefont {M.}~\bibnamefont
  {Avdeev}}, \ and\ \bibinfo {author} {\bibfnamefont {A.~J.}\ \bibnamefont
  {Studer}},\ }\href@noop {} {\bibfield  {journal} {\bibinfo  {journal} {Phys.
  Rev. B}\ }\textbf {\bibinfo {volume} {82}},\ \bibinfo {pages} {104502}
  (\bibinfo {year} {2010})}\BibitemShut {NoStop}%
\bibitem [{\citenamefont {Williams}\ \emph {et~al.}(2009)\citenamefont
  {Williams}, \citenamefont {McQueen}, \citenamefont {Ksenofontov},
  \citenamefont {Felser},\ and\ \citenamefont {Cava}}]{williams-2009-21}%
  \BibitemOpen
  \bibfield  {author} {\bibinfo {author} {\bibfnamefont {A.~J.}\ \bibnamefont
  {Williams}}, \bibinfo {author} {\bibfnamefont {T.~M.}\ \bibnamefont
  {McQueen}}, \bibinfo {author} {\bibfnamefont {V.}~\bibnamefont
  {Ksenofontov}}, \bibinfo {author} {\bibfnamefont {C.}~\bibnamefont {Felser}},
  \ and\ \bibinfo {author} {\bibfnamefont {R.~J.}\ \bibnamefont {Cava}},\
  }\href@noop {} {\bibfield  {journal} {\bibinfo  {journal} {J. Phys. Condens.
  Matter}\ }\textbf {\bibinfo {volume} {21}},\ \bibinfo {pages} {305701}
  (\bibinfo {year} {2009})}\BibitemShut {NoStop}%
\bibitem [{\citenamefont {Xu}\ \emph {et~al.}(2012)\citenamefont {Xu},
  \citenamefont {Wen}, \citenamefont {Zhao}, \citenamefont {Matsuda},
  \citenamefont {Ku}, \citenamefont {Liu}, \citenamefont {Gu}, \citenamefont
  {Lee}, \citenamefont {Birgeneau}, \citenamefont {Tranquada},\ and\
  \citenamefont {Xu}}]{PhysRevLett.109.227002}%
  \BibitemOpen
  \bibfield  {author} {\bibinfo {author} {\bibfnamefont {Z.}~\bibnamefont
  {Xu}}, \bibinfo {author} {\bibfnamefont {J.}~\bibnamefont {Wen}}, \bibinfo
  {author} {\bibfnamefont {Y.}~\bibnamefont {Zhao}}, \bibinfo {author}
  {\bibfnamefont {M.}~\bibnamefont {Matsuda}}, \bibinfo {author} {\bibfnamefont
  {W.}~\bibnamefont {Ku}}, \bibinfo {author} {\bibfnamefont {X.}~\bibnamefont
  {Liu}}, \bibinfo {author} {\bibfnamefont {G.}~\bibnamefont {Gu}}, \bibinfo
  {author} {\bibfnamefont {D.-H.}\ \bibnamefont {Lee}}, \bibinfo {author}
  {\bibfnamefont {R.~J.}\ \bibnamefont {Birgeneau}}, \bibinfo {author}
  {\bibfnamefont {J.~M.}\ \bibnamefont {Tranquada}}, \ and\ \bibinfo {author}
  {\bibfnamefont {G.}~\bibnamefont {Xu}},\ }\href {\doibase
  10.1103/PhysRevLett.109.227002} {\bibfield  {journal} {\bibinfo  {journal}
  {Phys. Rev. Lett.}\ }\textbf {\bibinfo {volume} {109}},\ \bibinfo {pages}
  {227002} (\bibinfo {year} {2012})}\BibitemShut {NoStop}%
\bibitem [{\citenamefont {Tsyrulin}\ \emph {et~al.}(2012)\citenamefont
  {Tsyrulin}, \citenamefont {Viennois}, \citenamefont {Giannini}, \citenamefont
  {Boehm}, \citenamefont {Jimenez-Ruiz}, \citenamefont {Omrani}, \citenamefont
  {Piazza},\ and\ \citenamefont {R\o{}nnow}}]{1367-2630-14-7-073025}%
  \BibitemOpen
  \bibfield  {author} {\bibinfo {author} {\bibfnamefont {N.}~\bibnamefont
  {Tsyrulin}}, \bibinfo {author} {\bibfnamefont {R.}~\bibnamefont {Viennois}},
  \bibinfo {author} {\bibfnamefont {E.}~\bibnamefont {Giannini}}, \bibinfo
  {author} {\bibfnamefont {M.}~\bibnamefont {Boehm}}, \bibinfo {author}
  {\bibfnamefont {M.}~\bibnamefont {Jimenez-Ruiz}}, \bibinfo {author}
  {\bibfnamefont {A.~A.}\ \bibnamefont {Omrani}}, \bibinfo {author}
  {\bibfnamefont {B.~D.}\ \bibnamefont {Piazza}}, \ and\ \bibinfo {author}
  {\bibfnamefont {H.~M.}\ \bibnamefont {R\o{}nnow}},\ }\href@noop {} {\bibfield
   {journal} {\bibinfo  {journal} {New J. Phys.}\ }\textbf {\bibinfo {volume}
  {14}},\ \bibinfo {pages} {073025} (\bibinfo {year} {2012})}\BibitemShut
  {NoStop}%
\bibitem [{\citenamefont {Lumsden}\ \emph {et~al.}(2010)\citenamefont
  {Lumsden}, \citenamefont {Christianson}, \citenamefont {Goremychkin},
  \citenamefont {Nagler}, \citenamefont {Mook}, \citenamefont {Stone},
  \citenamefont {Abernathy}, \citenamefont {Guidi}, \citenamefont {MacDougall},
  \citenamefont {{De La Cruz}}, \citenamefont {Sefat}, \citenamefont {McGuire},
  \citenamefont {Sales},\ and\ \citenamefont {Mandrus}}]{lumsden-2009}%
  \BibitemOpen
  \bibfield  {author} {\bibinfo {author} {\bibfnamefont {M.~D.}\ \bibnamefont
  {Lumsden}}, \bibinfo {author} {\bibfnamefont {A.~D.}\ \bibnamefont
  {Christianson}}, \bibinfo {author} {\bibfnamefont {E.~A.}\ \bibnamefont
  {Goremychkin}}, \bibinfo {author} {\bibfnamefont {S.~E.}\ \bibnamefont
  {Nagler}}, \bibinfo {author} {\bibfnamefont {H.~A.}\ \bibnamefont {Mook}},
  \bibinfo {author} {\bibfnamefont {M.~B.}\ \bibnamefont {Stone}}, \bibinfo
  {author} {\bibfnamefont {D.~L.}\ \bibnamefont {Abernathy}}, \bibinfo {author}
  {\bibfnamefont {T.}~\bibnamefont {Guidi}}, \bibinfo {author} {\bibfnamefont
  {G.~J.}\ \bibnamefont {MacDougall}}, \bibinfo {author} {\bibfnamefont
  {C.}~\bibnamefont {{De La Cruz}}}, \bibinfo {author} {\bibfnamefont {A.~S.}\
  \bibnamefont {Sefat}}, \bibinfo {author} {\bibfnamefont {M.~A.}\ \bibnamefont
  {McGuire}}, \bibinfo {author} {\bibfnamefont {B.~C.}\ \bibnamefont {Sales}},
  \ and\ \bibinfo {author} {\bibfnamefont {D.}~\bibnamefont {Mandrus}},\
  }\href@noop {} {\bibfield  {journal} {\bibinfo  {journal} {Nature Phys.}\
  }\textbf {\bibinfo {volume} {6}},\ \bibinfo {pages} {182} (\bibinfo {year}
  {2010})}\BibitemShut {NoStop}%
\bibitem [{\citenamefont {{Liu}}\ \emph {et~al.}(2010)\citenamefont {{Liu}},
  \citenamefont {{Hu}}, \citenamefont {{Qian}}, \citenamefont {{Fobes}},
  \citenamefont {{Mao}}, \citenamefont {{Bao}}, \citenamefont {{Reehuis}},
  \citenamefont {{Kimber}}, \citenamefont {{Prokes}}, \citenamefont {{Matas}},
  \citenamefont {{Argyriou}}, \citenamefont {{Hiess}}, \citenamefont
  {{Rotaru}}, \citenamefont {{Pham}}, \citenamefont {{Spinu}}, \citenamefont
  {{Qiu}}, \citenamefont {{Thampy}}, \citenamefont {{Savici}}, \citenamefont
  {{Rodriguez}},\ and\ \citenamefont {{Broholm}}}]{liupi0topp}%
  \BibitemOpen
  \bibfield  {author} {\bibinfo {author} {\bibfnamefont {T.~J.}\ \bibnamefont
  {{Liu}}}, \bibinfo {author} {\bibfnamefont {J.}~\bibnamefont {{Hu}}},
  \bibinfo {author} {\bibfnamefont {B.}~\bibnamefont {{Qian}}}, \bibinfo
  {author} {\bibfnamefont {D.}~\bibnamefont {{Fobes}}}, \bibinfo {author}
  {\bibfnamefont {Z.~Q.}\ \bibnamefont {{Mao}}}, \bibinfo {author}
  {\bibfnamefont {W.}~\bibnamefont {{Bao}}}, \bibinfo {author} {\bibfnamefont
  {M.}~\bibnamefont {{Reehuis}}}, \bibinfo {author} {\bibfnamefont {S.~A.~J.}\
  \bibnamefont {{Kimber}}}, \bibinfo {author} {\bibfnamefont {K.}~\bibnamefont
  {{Prokes}}}, \bibinfo {author} {\bibfnamefont {S.}~\bibnamefont {{Matas}}},
  \bibinfo {author} {\bibfnamefont {D.~N.}\ \bibnamefont {{Argyriou}}},
  \bibinfo {author} {\bibfnamefont {A.}~\bibnamefont {{Hiess}}}, \bibinfo
  {author} {\bibfnamefont {A.}~\bibnamefont {{Rotaru}}}, \bibinfo {author}
  {\bibfnamefont {H.}~\bibnamefont {{Pham}}}, \bibinfo {author} {\bibfnamefont
  {L.}~\bibnamefont {{Spinu}}}, \bibinfo {author} {\bibfnamefont
  {Y.}~\bibnamefont {{Qiu}}}, \bibinfo {author} {\bibfnamefont
  {V.}~\bibnamefont {{Thampy}}}, \bibinfo {author} {\bibfnamefont {A.~T.}\
  \bibnamefont {{Savici}}}, \bibinfo {author} {\bibfnamefont {J.~A.}\
  \bibnamefont {{Rodriguez}}}, \ and\ \bibinfo {author} {\bibfnamefont
  {C.}~\bibnamefont {{Broholm}}},\ }\href@noop {} {\bibfield  {journal}
  {\bibinfo  {journal} {Nature Mater.}\ }\textbf {\bibinfo {volume} {9}},\
  \bibinfo {pages} {718} (\bibinfo {year} {2010})}\BibitemShut {NoStop}%
\bibitem [{\citenamefont {Christianson}\ \emph {et~al.}(2013)\citenamefont
  {Christianson}, \citenamefont {Lumsden}, \citenamefont {Marty}, \citenamefont
  {Wang}, \citenamefont {Calder}, \citenamefont {Abernathy}, \citenamefont
  {Stone}, \citenamefont {Mook}, \citenamefont {McGuire}, \citenamefont
  {Sefat}, \citenamefont {Sales}, \citenamefont {Mandrus},\ and\ \citenamefont
  {Goremychkin}}]{PhysRevB.87.224410}%
  \BibitemOpen
  \bibfield  {author} {\bibinfo {author} {\bibfnamefont {A.~D.}\ \bibnamefont
  {Christianson}}, \bibinfo {author} {\bibfnamefont {M.~D.}\ \bibnamefont
  {Lumsden}}, \bibinfo {author} {\bibfnamefont {K.}~\bibnamefont {Marty}},
  \bibinfo {author} {\bibfnamefont {C.~H.}\ \bibnamefont {Wang}}, \bibinfo
  {author} {\bibfnamefont {S.}~\bibnamefont {Calder}}, \bibinfo {author}
  {\bibfnamefont {D.~L.}\ \bibnamefont {Abernathy}}, \bibinfo {author}
  {\bibfnamefont {M.~B.}\ \bibnamefont {Stone}}, \bibinfo {author}
  {\bibfnamefont {H.~A.}\ \bibnamefont {Mook}}, \bibinfo {author}
  {\bibfnamefont {M.~A.}\ \bibnamefont {McGuire}}, \bibinfo {author}
  {\bibfnamefont {A.~S.}\ \bibnamefont {Sefat}}, \bibinfo {author}
  {\bibfnamefont {B.~C.}\ \bibnamefont {Sales}}, \bibinfo {author}
  {\bibfnamefont {D.}~\bibnamefont {Mandrus}}, \ and\ \bibinfo {author}
  {\bibfnamefont {E.~A.}\ \bibnamefont {Goremychkin}},\ }\href {\doibase
  10.1103/PhysRevB.87.224410} {\bibfield  {journal} {\bibinfo  {journal} {Phys.
  Rev. B}\ }\textbf {\bibinfo {volume} {87}},\ \bibinfo {pages} {224410}
  (\bibinfo {year} {2013})}\BibitemShut {NoStop}%
\bibitem [{\citenamefont {Yin}\ \emph {et~al.}(2010)\citenamefont {Yin},
  \citenamefont {Lee},\ and\ \citenamefont {Ku}}]{weiguounified}%
  \BibitemOpen
  \bibfield  {author} {\bibinfo {author} {\bibfnamefont {W.-G.}\ \bibnamefont
  {Yin}}, \bibinfo {author} {\bibfnamefont {C.-C.}\ \bibnamefont {Lee}}, \ and\
  \bibinfo {author} {\bibfnamefont {W.}~\bibnamefont {Ku}},\ }\href@noop {}
  {\bibfield  {journal} {\bibinfo  {journal} {Phys. Rev. Lett.}\ }\textbf
  {\bibinfo {volume} {105}},\ \bibinfo {pages} {107004} (\bibinfo {year}
  {2010})}\BibitemShut {NoStop}%
\bibitem [{\citenamefont {Johannes}\ and\ \citenamefont
  {Mazin}(2009)}]{johannes-2009}%
  \BibitemOpen
  \bibfield  {author} {\bibinfo {author} {\bibfnamefont {M.~D.}\ \bibnamefont
  {Johannes}}\ and\ \bibinfo {author} {\bibfnamefont {I.~I.}\ \bibnamefont
  {Mazin}},\ }\href@noop {} {\bibfield  {journal} {\bibinfo  {journal} {Phys.
  Rev. B}\ }\textbf {\bibinfo {volume} {79}},\ \bibinfo {pages} {220510(R)}
  (\bibinfo {year} {2009})}\BibitemShut {NoStop}%
\bibitem [{\citenamefont {Dai}\ \emph {et~al.}(2012)\citenamefont {Dai},
  \citenamefont {Hu},\ and\ \citenamefont {Dagotto}}]{np_8_709}%
  \BibitemOpen
  \bibfield  {author} {\bibinfo {author} {\bibfnamefont {P.}~\bibnamefont
  {Dai}}, \bibinfo {author} {\bibfnamefont {J.}~\bibnamefont {Hu}}, \ and\
  \bibinfo {author} {\bibfnamefont {E.}~\bibnamefont {Dagotto}},\ }\href
  {\doibase 10.1038/nphys2438} {\bibfield  {journal} {\bibinfo  {journal} {Nat
  Phys}\ }\textbf {\bibinfo {volume} {8}},\ \bibinfo {pages} {709} (\bibinfo
  {year} {2012})}\BibitemShut {NoStop}%
\bibitem [{\citenamefont {Liu}\ \emph {et~al.}(2012)\citenamefont {Liu},
  \citenamefont {Harriger}, \citenamefont {Luo}, \citenamefont {Wang},
  \citenamefont {Ewings}, \citenamefont {Guidi}, \citenamefont {Park},
  \citenamefont {Haule}, \citenamefont {Kotliar}, \citenamefont {Hayden},\ and\
  \citenamefont {Dai}}]{arXiv:1202.2827}%
  \BibitemOpen
  \bibfield  {author} {\bibinfo {author} {\bibfnamefont {M.}~\bibnamefont
  {Liu}}, \bibinfo {author} {\bibfnamefont {L.~W.}\ \bibnamefont {Harriger}},
  \bibinfo {author} {\bibfnamefont {H.}~\bibnamefont {Luo}}, \bibinfo {author}
  {\bibfnamefont {M.}~\bibnamefont {Wang}}, \bibinfo {author} {\bibfnamefont
  {R.~A.}\ \bibnamefont {Ewings}}, \bibinfo {author} {\bibfnamefont
  {T.}~\bibnamefont {Guidi}}, \bibinfo {author} {\bibfnamefont
  {H.}~\bibnamefont {Park}}, \bibinfo {author} {\bibfnamefont {K.}~\bibnamefont
  {Haule}}, \bibinfo {author} {\bibfnamefont {G.}~\bibnamefont {Kotliar}},
  \bibinfo {author} {\bibfnamefont {S.~M.}\ \bibnamefont {Hayden}}, \ and\
  \bibinfo {author} {\bibfnamefont {P.}~\bibnamefont {Dai}},\ }\href {\doibase
  10.1038/nphys2268} {\bibfield  {journal} {\bibinfo  {journal} {Nat Phys}\
  }\textbf {\bibinfo {volume} {8}},\ \bibinfo {pages} {376} (\bibinfo {year}
  {2012})}\BibitemShut {NoStop}%
\bibitem [{\citenamefont {Gretarsson}\ \emph {et~al.}(2011)\citenamefont
  {Gretarsson}, \citenamefont {Lupascu}, \citenamefont {Kim}, \citenamefont
  {Casa}, \citenamefont {Gog}, \citenamefont {Wu}, \citenamefont {Julian},
  \citenamefont {Xu}, \citenamefont {Wen}, \citenamefont {Gu}, \citenamefont
  {Yuan}, \citenamefont {Chen}, \citenamefont {Wang}, \citenamefont {Khim},
  \citenamefont {Kim}, \citenamefont {Ishikado}, \citenamefont {Jarrige},
  \citenamefont {Shamoto}, \citenamefont {Chu}, \citenamefont {Fisher},\ and\
  \citenamefont {Kim}}]{PhysRevB.84.100509}%
  \BibitemOpen
  \bibfield  {author} {\bibinfo {author} {\bibfnamefont {H.}~\bibnamefont
  {Gretarsson}}, \bibinfo {author} {\bibfnamefont {A.}~\bibnamefont {Lupascu}},
  \bibinfo {author} {\bibfnamefont {J.}~\bibnamefont {Kim}}, \bibinfo {author}
  {\bibfnamefont {D.}~\bibnamefont {Casa}}, \bibinfo {author} {\bibfnamefont
  {T.}~\bibnamefont {Gog}}, \bibinfo {author} {\bibfnamefont {W.}~\bibnamefont
  {Wu}}, \bibinfo {author} {\bibfnamefont {S.~R.}\ \bibnamefont {Julian}},
  \bibinfo {author} {\bibfnamefont {Z.~J.}\ \bibnamefont {Xu}}, \bibinfo
  {author} {\bibfnamefont {J.~S.}\ \bibnamefont {Wen}}, \bibinfo {author}
  {\bibfnamefont {G.~D.}\ \bibnamefont {Gu}}, \bibinfo {author} {\bibfnamefont
  {R.~H.}\ \bibnamefont {Yuan}}, \bibinfo {author} {\bibfnamefont {Z.~G.}\
  \bibnamefont {Chen}}, \bibinfo {author} {\bibfnamefont {N.-L.}\ \bibnamefont
  {Wang}}, \bibinfo {author} {\bibfnamefont {S.}~\bibnamefont {Khim}}, \bibinfo
  {author} {\bibfnamefont {K.~H.}\ \bibnamefont {Kim}}, \bibinfo {author}
  {\bibfnamefont {M.}~\bibnamefont {Ishikado}}, \bibinfo {author}
  {\bibfnamefont {I.}~\bibnamefont {Jarrige}}, \bibinfo {author} {\bibfnamefont
  {S.}~\bibnamefont {Shamoto}}, \bibinfo {author} {\bibfnamefont {J.-H.}\
  \bibnamefont {Chu}}, \bibinfo {author} {\bibfnamefont {I.~R.}\ \bibnamefont
  {Fisher}}, \ and\ \bibinfo {author} {\bibfnamefont {Y.-J.}\ \bibnamefont
  {Kim}},\ }\href {\doibase 10.1103/PhysRevB.84.100509} {\bibfield  {journal}
  {\bibinfo  {journal} {Phys. Rev. B}\ }\textbf {\bibinfo {volume} {84}},\
  \bibinfo {pages} {100509} (\bibinfo {year} {2011})}\BibitemShut {NoStop}%
\bibitem [{\citenamefont {Kou}\ \emph {et~al.}(2009)\citenamefont {Kou},
  \citenamefont {Li},\ and\ \citenamefont {Weng}}]{0295-5075-88-1-17010}%
  \BibitemOpen
  \bibfield  {author} {\bibinfo {author} {\bibfnamefont {S.-P.}\ \bibnamefont
  {Kou}}, \bibinfo {author} {\bibfnamefont {T.}~\bibnamefont {Li}}, \ and\
  \bibinfo {author} {\bibfnamefont {Z.-Y.}\ \bibnamefont {Weng}},\ }\href
  {http://stacks.iop.org/0295-5075/88/i=1/a=17010} {\bibfield  {journal}
  {\bibinfo  {journal} {Euro. Phys. Lett.}\ }\textbf {\bibinfo {volume} {88}},\
  \bibinfo {pages} {17010} (\bibinfo {year} {2009})}\BibitemShut {NoStop}%
\bibitem [{\citenamefont {Zaliznyak}\ \emph {et~al.}(2011)\citenamefont
  {Zaliznyak}, \citenamefont {Xu}, \citenamefont {Tranquada}, \citenamefont
  {Gu}, \citenamefont {Tsvelik},\ and\ \citenamefont
  {Stone}}]{2011arXiv1103.5073Z}%
  \BibitemOpen
  \bibfield  {author} {\bibinfo {author} {\bibfnamefont {I.~A.}\ \bibnamefont
  {Zaliznyak}}, \bibinfo {author} {\bibfnamefont {Z.}~\bibnamefont {Xu}},
  \bibinfo {author} {\bibfnamefont {J.~M.}\ \bibnamefont {Tranquada}}, \bibinfo
  {author} {\bibfnamefont {G.}~\bibnamefont {Gu}}, \bibinfo {author}
  {\bibfnamefont {A.~M.}\ \bibnamefont {Tsvelik}}, \ and\ \bibinfo {author}
  {\bibfnamefont {M.~B.}\ \bibnamefont {Stone}},\ }\href@noop {} {\bibfield
  {journal} {\bibinfo  {journal} {Phys. Rev. Lett.}\ }\textbf {\bibinfo
  {volume} {107}},\ \bibinfo {pages} {216403} (\bibinfo {year}
  {2011})}\BibitemShut {NoStop}%
\bibitem [{\citenamefont {Chadov}\ \emph {et~al.}(2010)\citenamefont {Chadov},
  \citenamefont {Sch\"arf}, \citenamefont {Fecher}, \citenamefont {Felser},
  \citenamefont {Zhang},\ and\ \citenamefont {Singh}}]{localm3}%
  \BibitemOpen
  \bibfield  {author} {\bibinfo {author} {\bibfnamefont {S.}~\bibnamefont
  {Chadov}}, \bibinfo {author} {\bibfnamefont {D.}~\bibnamefont {Sch\"arf}},
  \bibinfo {author} {\bibfnamefont {G.~H.}\ \bibnamefont {Fecher}}, \bibinfo
  {author} {\bibfnamefont {C.}~\bibnamefont {Felser}}, \bibinfo {author}
  {\bibfnamefont {L.}~\bibnamefont {Zhang}}, \ and\ \bibinfo {author}
  {\bibfnamefont {D.~J.}\ \bibnamefont {Singh}},\ }\href@noop {} {\bibfield
  {journal} {\bibinfo  {journal} {Phys. Rev. B}\ }\textbf {\bibinfo {volume}
  {81}},\ \bibinfo {pages} {104523} (\bibinfo {year} {2010})}\BibitemShut
  {NoStop}%
\bibitem [{\citenamefont {McLeod}\ \emph {et~al.}(2012)\citenamefont {McLeod},
  \citenamefont {Buling}, \citenamefont {Green}, \citenamefont {Boyko},
  \citenamefont {Skorikov}, \citenamefont {Kurmaev}, \citenamefont {Neumann},
  \citenamefont {Finkelstein}, \citenamefont {Ni}, \citenamefont {Thaler},
  \citenamefont {Bud?ko}, \citenamefont {Canfield},\ and\ \citenamefont
  {Moewes}}]{0953-8984-24-21-215501}%
  \BibitemOpen
  \bibfield  {author} {\bibinfo {author} {\bibfnamefont {J.~A.}\ \bibnamefont
  {McLeod}}, \bibinfo {author} {\bibfnamefont {A.}~\bibnamefont {Buling}},
  \bibinfo {author} {\bibfnamefont {R.~J.}\ \bibnamefont {Green}}, \bibinfo
  {author} {\bibfnamefont {T.~D.}\ \bibnamefont {Boyko}}, \bibinfo {author}
  {\bibfnamefont {N.~A.}\ \bibnamefont {Skorikov}}, \bibinfo {author}
  {\bibfnamefont {E.~Z.}\ \bibnamefont {Kurmaev}}, \bibinfo {author}
  {\bibfnamefont {M.}~\bibnamefont {Neumann}}, \bibinfo {author} {\bibfnamefont
  {L.~D.}\ \bibnamefont {Finkelstein}}, \bibinfo {author} {\bibfnamefont
  {N.}~\bibnamefont {Ni}}, \bibinfo {author} {\bibfnamefont {A.}~\bibnamefont
  {Thaler}}, \bibinfo {author} {\bibfnamefont {S.~L.}\ \bibnamefont {Bud?ko}},
  \bibinfo {author} {\bibfnamefont {P.~C.}\ \bibnamefont {Canfield}}, \ and\
  \bibinfo {author} {\bibfnamefont {A.}~\bibnamefont {Moewes}},\ }\href
  {http://stacks.iop.org/0953-8984/24/i=21/a=215501} {\bibfield  {journal}
  {\bibinfo  {journal} {Journal of Physics: Condensed Matter}\ }\textbf
  {\bibinfo {volume} {24}},\ \bibinfo {pages} {215501} (\bibinfo {year}
  {2012})}\BibitemShut {NoStop}%
\end{thebibliography}
%

\end{document}